\documentclass[12pt,manuscript]{aastex}
\usepackage{emulateapj5}
\usepackage{psfig}

\def\msun{M$_{\odot}$}
\def\eg{{\em eg.\ }}

\def\lesssim{\mathrel{\hbox{\rlap{\hbox{\lower4pt\hbox{$\sim$}}}\hbox{$<$}}}}
\def\gtrsim{\mathrel{\hbox{\rlap{\hbox{\lower4pt\hbox{$\sim$}}}\hbox{$>$}}}}

\begin{document}

\title{Constraints on Galaxy Formation from Stars in the Far Outer Disk of M31\footnotemark[1]}

\footnotetext[1]{Based on observations
with the NASA/ESA Hubble Space Telescope, obtained from the 
data archive of the Space
Telescope Science Institute.  STScI is operated by the Association of
Universities for Research in Astronomy, Inc. under NASA contract No.
NAS5-26555.}

\author{Annette M.~N. Ferguson}
\affil{Kapteyn Astronomical Institute, Postbus 800, 9700 AV Groningen, The Netherlands}
\email{ferguson@astro.rug.nl}

\author{Rachel A. Johnson}
\affil{Institute of Astronomy, University of Cambridge, Madingley Road,\\
Cambridge, UK CB3 0HA}
\email{raj@ast.cam.ac.uk}

\begin{abstract}
Numerical simulations of galaxy formation within the cold dark matter
(CDM) hierarchical clustering framework are unable to produce large
disk galaxies without invoking some form of feedback to suppress gas
cooling and collapse until a redshift of unity or below.  An important
observational consequence of delaying the epoch of disk  formation
until relatively recent times is that the stellar populations in the
extended disk should of be predominantly young-to-intermediate age.  We
use a deep HST/WFPC2 archival pointing to investigate the mean age and
metallicity of the stellar population in a disk-dominated field at
30~kpc along the major axis of M31.  Our analysis of the
color-magnitude-diagram reveals the dominant population to have
significant mean age ($\gtrsim 8$Gyr) and a moderately-high mean
metallicity ([Fe/H]$\sim -0.7$); tentative evidence is also presented for a
trace population of ancient ($\ge 10$ Gyr) metal-poor stars.
These characteristics are unexpected in CDM models and we discuss the
possible implications of this result, as well as alternative
interpretations.
\end{abstract}

\section{Introduction}

It is generally believed that galactic disks form from baryons which
cool and dissipatively collapse inside the potential wells of
tidally-torqued dark matter halos.  Under the assumption that the gas
component retains most of its initial angular momentum, this process
leads to systems with sizes and collapse times compatible with
present-day large disk galaxies (\eg  Fall \& Efstathiou 1980).
However, numerical simulations of galaxy formation within the popular
cold dark matter (CDM) hierarchical clustering  framework indicate that
this condition is far from being satisfied.  The merging process
inherent in this picture leads to  efficient outward transport of
angular momentum from the collapsing gas to the dark halo, resulting in
final disks which have specific angular momenta an order of magnitude
too small (\eg Navarro \& Steinmetz 1997).

A possible solution to this ``angular momemtum problem" is to suppress
gas cooling and collapse  until late times, when the most active phase
of merging is over.    Various papers have shown that if strong
feedback from an early generation of stars can delay the epoch of disk
formation to z$\lesssim1$ then systems  can form by the present-day with
angular momenta compatible with observations (\eg Weil, Eke \&
Efstathiou 1998; Sommer-Larsen, Gelato \& Vedel 1999; Binney, Gerhard
\& Silk 2001).

While delayed disk formation is a promising solution
to the angular momentum problem, one must ask whether such a scenario
is compatible with the age distribution of stars in present-day disks.
For a flat Universe with $\Omega_{\Lambda}=1-{\Omega_{matter}}=0.7$ and
H$_o=65-75~\rm{km}~{s^{-1}}~Mpc^{-1}$, a redshift of unity corresponds
to a lookback-time of 7--8~Gyr.   Binney, Dehnen \& Bertelli (2000)
recently determined the ages of the oldest thin disk stars in the
Hipparcos sample to be $\approx$11~Gyr, on a scale where the halo
globular clusters are 12~Gyr old, although they state that maximum ages
as low as 9~Gyr are plausible.  Adopting this as a lower limit
to the age of the Milky Way disk, one infers that at least some
fraction of the local thin disk -- corresponding to 8 kpc in radius, or
$\sim 3$ exponential scalelengths -- was in place by a redshift of
$\sim$1.4 for a flat Universe cosmology.  Extending these arguments to
the local thick disk could push this redshift back to  $\sim 2$ (\eg
Wyse 2001).

While the local Galactic disk places an interesting constraint on the
epoch of disk formation, it is clearly desirable to explore other
localities and other galaxies.   Indeed, the entire disk may not
necessarily need to form late in delayed formation models; it is
possible that a less extended disk  was in place at early times, with
the accretion of higher angular momentum material at later epochs (\eg
Ferguson \& Clarke 2001).  A knowledge of the stellar
 populations at very large radii in galactic disks is therefore of
particular importance.

This {\it Letter} presents the first  results from a study to probe the
fossil record in our nearest large neighbour, M31, through the analysis
of deep, archival HST/WFPC2 pointings. We focus here on the
color-magnitude-diagram (CMD) of a disk-dominated field in the far
outer disk of M31 which proves to have potentially interesting
implications for the formation epoch of large disk galaxies.

\section{Observations \& Photometric Reduction}

The M31 globular cluster G327 was targeted with HST/WFPC2 as part of
program GO6671.  Exposure times of 5300 and 5400s were obtained in filters
F555W and F814W, respectively.  Due to an error in
coordinates, the actual telescope pointing (PC centered on
$\alpha_{2000}=00{^h}49{^m}36.2{^s},
\delta_{2000}=43{\arcdeg}01{\arcmin}07{\arcsec})$ was some distance
from G327 and placed the field along the north-west major axis at a
radial distance of $\sim$30~kpc, where we assume $m-M=24.47 \pm 0.12$
(D=783~kpc; Durrell et al 2001).
 For M31 position angles in the range 35-40{\arcdeg}, the WFPC2 field
lies $\lesssim$ 5{\arcdeg} from the major axis (see Figure 1).  By
extrapolating the structural parameters determined by Walterbos \&
Kennicutt (1988, hereafter WK88) and assuming an inclination of
12.5\arcdeg, we expect that at this location -- corresponding to
approximately 5 exponential disk R-band scalelengths or 1.4~R$_{25}$ --
disk stars should contribute $\sim 95\%$ of the stellar surface
density.

\centerline{\psfig{file=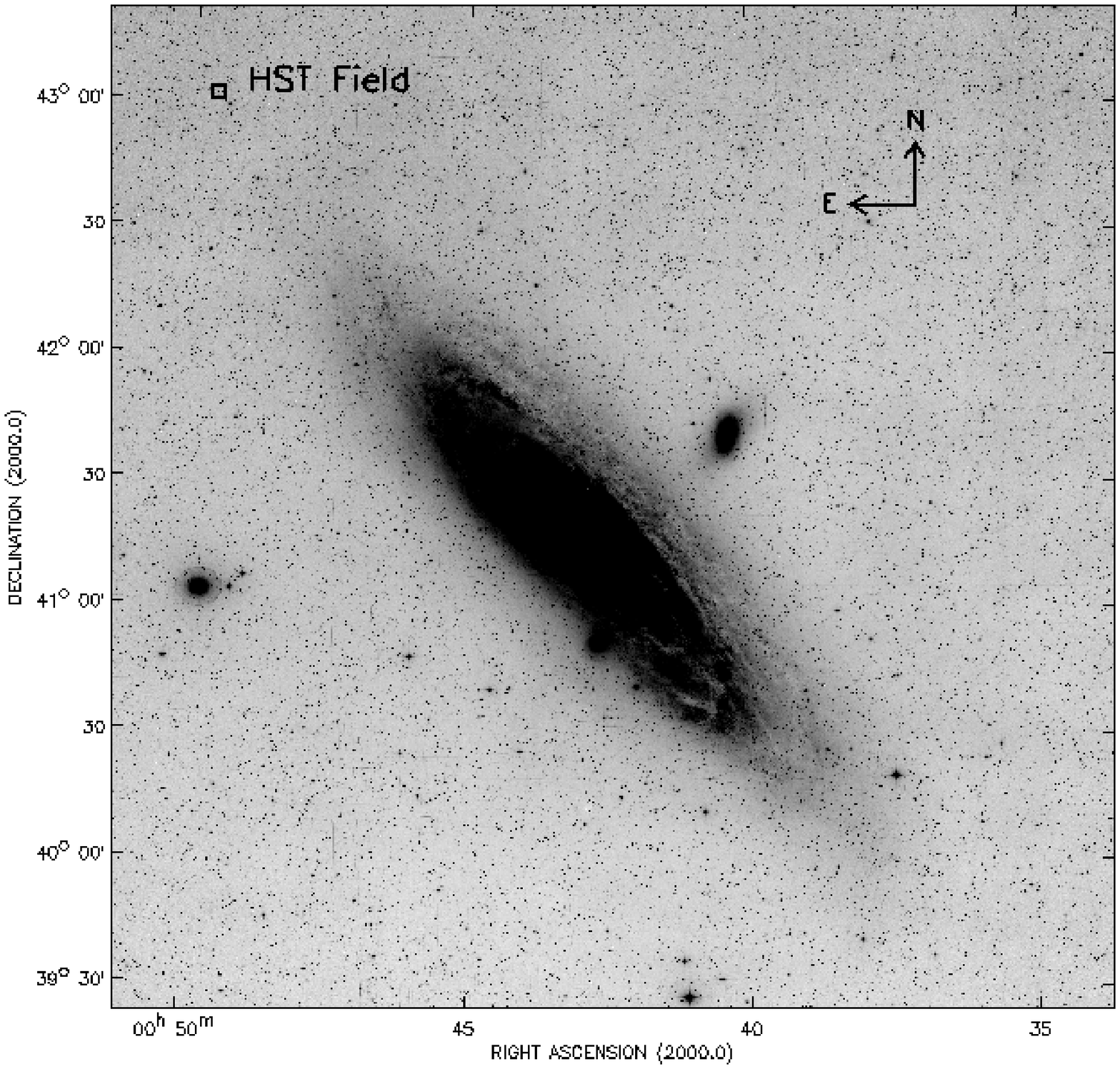,height=7cm}}
{\small{\bf Fig.1 - } A Palomar Sky Survey plate of M31 with the
location of the HST field overlaid. The field size is $3.4\arcdeg
\times 4.0\arcdeg$; the box indicating the WFPC2 pointing is not drawn
to scale.}
\bigskip

The data were retrieved from the Space Telescope-European Coordinating
Facility archive and reprocessed with the most up-to-date calibration
exposures.  Due to the relative sparseness of the field, only the
larger-area WF frames were considered. Images
in a given filter were combined with a cosmic-ray rejection algorithm
and Tiny Tim point-spread functions (PSFs) were fitted to the detected
stars using the DAOPHOT/ALLSTAR crowded-field photometry package
(Stetson 1987). Stars detected in different filters were matched and
the photometry lists were pruned to exclude objects which were either
poorly fit by the PSF model or had large  (ie. $\sigma \gtrsim 0.3$
mag) photometric errors. Aperture corrections were measured from stars
on our frames and the synthetic transformations of Holtzman et al
(1995) were applied to derive standard magnitudes.  We adopt a
foreground Galactic reddening of E(V$-$I)$=$0.10 toward M31 (Holland
et al 1996); as the field lies a considerable distance from the center
of the galaxy (albeit still within the low density outer reaches of
the HI disk), no correction was made for internal reddening.  Artificial star
tests indicate a high level of completeness with $\gtrsim$ 75\% of the
stars returned for V $\sim$ 26.5, I $\sim$25.5.  Full details of the
analysis procedure will be reported elsewhere.

\section{Dissecting the Colour-Magnitude Diagram}

Figure 2 presents the (V,V$-$I) and (I,V$-$I) CMDs for the far outer
disk field.  The basic morphology is that of a predominantly
old-to-intermediate age population; indeed the CMDs bear a striking
resemblance to those of M31 {\it halo} fields studied previously with
HST (\eg Rich et al 1996, Holland et al 1996) however the halo
component is expected to be only a very minor contributor at this
location.  Contamination from foreground Galactic stars and background
unresolved galaxies is expected to be negligible in WFPC2 fields along
the M31 sightline (see Holland et al 1996, Ferguson et al 2000).

\subsection{The Red Giant Branch}

The most prominent features in the CMD are the red giant branch (RGB),
and the red clump (RC) which is superposed on the RGB at V $\sim 25.3$,
I $\sim 24.3$.  Together, these features contain more than 95\% of the
total number of stars detected in the WFPC2 field above the 75\%
completeness level.  The mere existence of these features in the CMD
attests to the presence of a population(s) with age(s) in the range $\sim
2-10$ Gyr. 

\centerline{\psfig{file=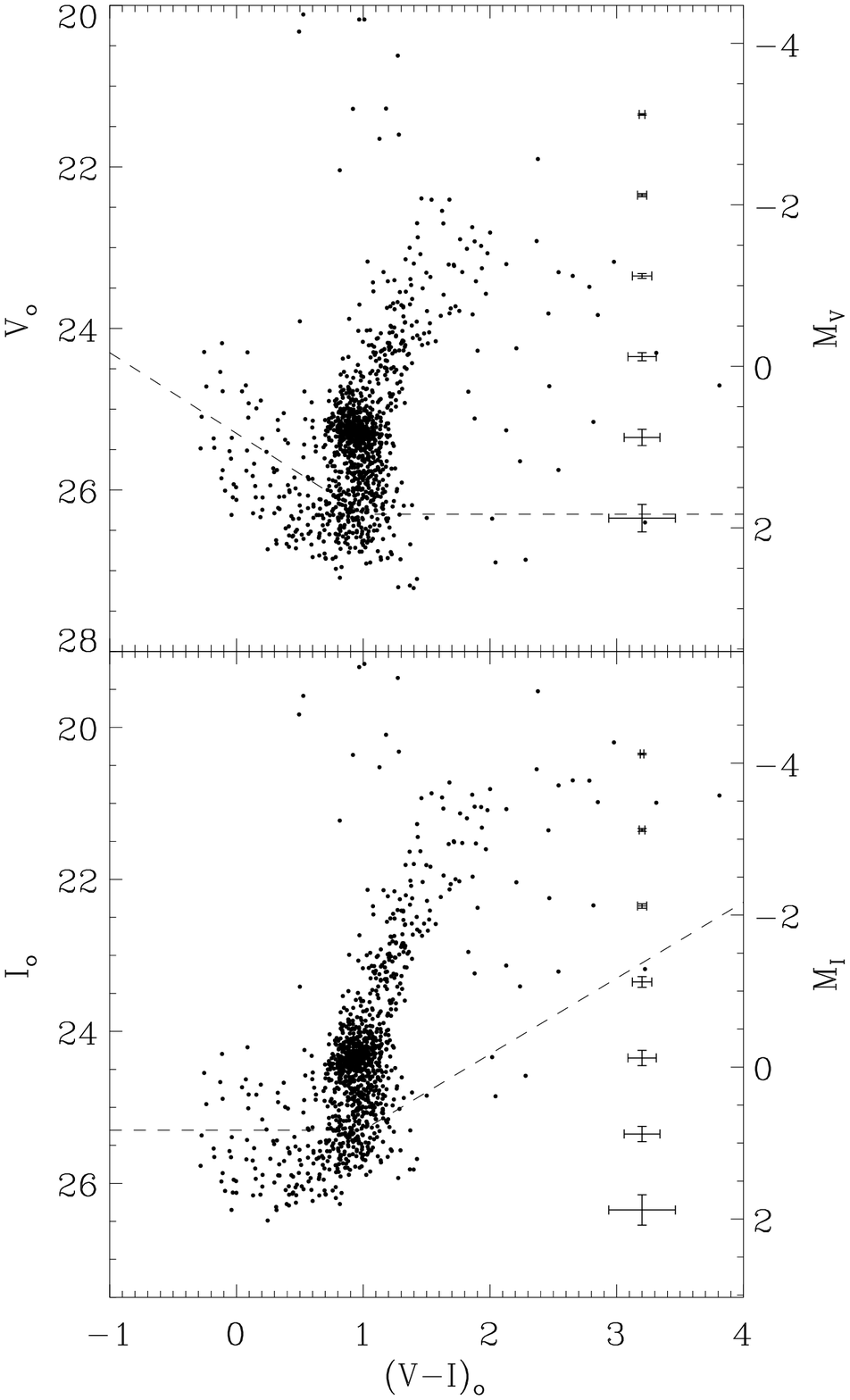,height=11cm}}
{\small{\bf Fig.2 - } The (V,V$-$I) (top) and (I,V$-$I) (bottom) CMDs for
stars at large radius along the major axis of M31 on the standard
Johnsons-Cousins system. Stars from all three WF chips are plotted. The
dashed line indicates the 75\% completeness level as determined from
fake star tests.  A distance modulus of 24.47 and a reddening of E(V-I)=0.10
(Holland et al 1996) have been used to transform to absolute magnitudes
and unreddened colors.}
\bigskip

The RGB V-I color is rather insensitive to ages $\gtrsim 2$ Gyr, but is
more sensitive to metallicity, as demonstrated in Figure 3 which shows
fiducial RGBs for Galactic globular clusters overlaid on the (V,V$-$I)
CMD.  The M31 outer disk RGB is well-matched in the mean by the ridge
line of 47 Tuc, the prototypical metal-rich globular cluster with
[Fe/H]$=-0.71$ on the Zinn \& West (1984) scale.  Very few stars are
visible above the tip of the giant branch.   The significant color
width of the RGB detected here ($\sim 1$ mag at V $\sim 23$) exceeds
that of photometric errors, and is most easily explained by an
intrinsic spread in metallicity of the stellar population.  Assuming an
old, nearly coeval population, the comparison with globular cluster
fiducials indicates the metallicity spread could be as large as $\sim
2$ dex.  On the other hand, a
mono-metallicity population with [Fe/H]$=-0.7$ would, using
the Girardi et al (2000) isochrones, have a maximum RGB width of only
$\lesssim 0.4$ mag for an age range of 0.5--16 Gyr.   It therefore
appears that the relatively high metallicity and intrinsic dispersion
which has been previously shown to characterise the M31 field halo (\eg
Mould \& Kristian 1986; Holland et al 1996; Rich et al 1996; Durrell et
al 2001) also characterises the far outer disk.

\subsection{The Red Clump}

The color and luminosity of a  core He-burning star depends on its
age,  metallicity and He content.  Several recent papers have
highlighted the  potential power of the red clump (RC), when
coupled with independent metallicity estimates, as an age indicator
(\eg Cole 1999; Cole et al 1999; Girardi \& Salaris 2001).  

\centerline{\psfig{file=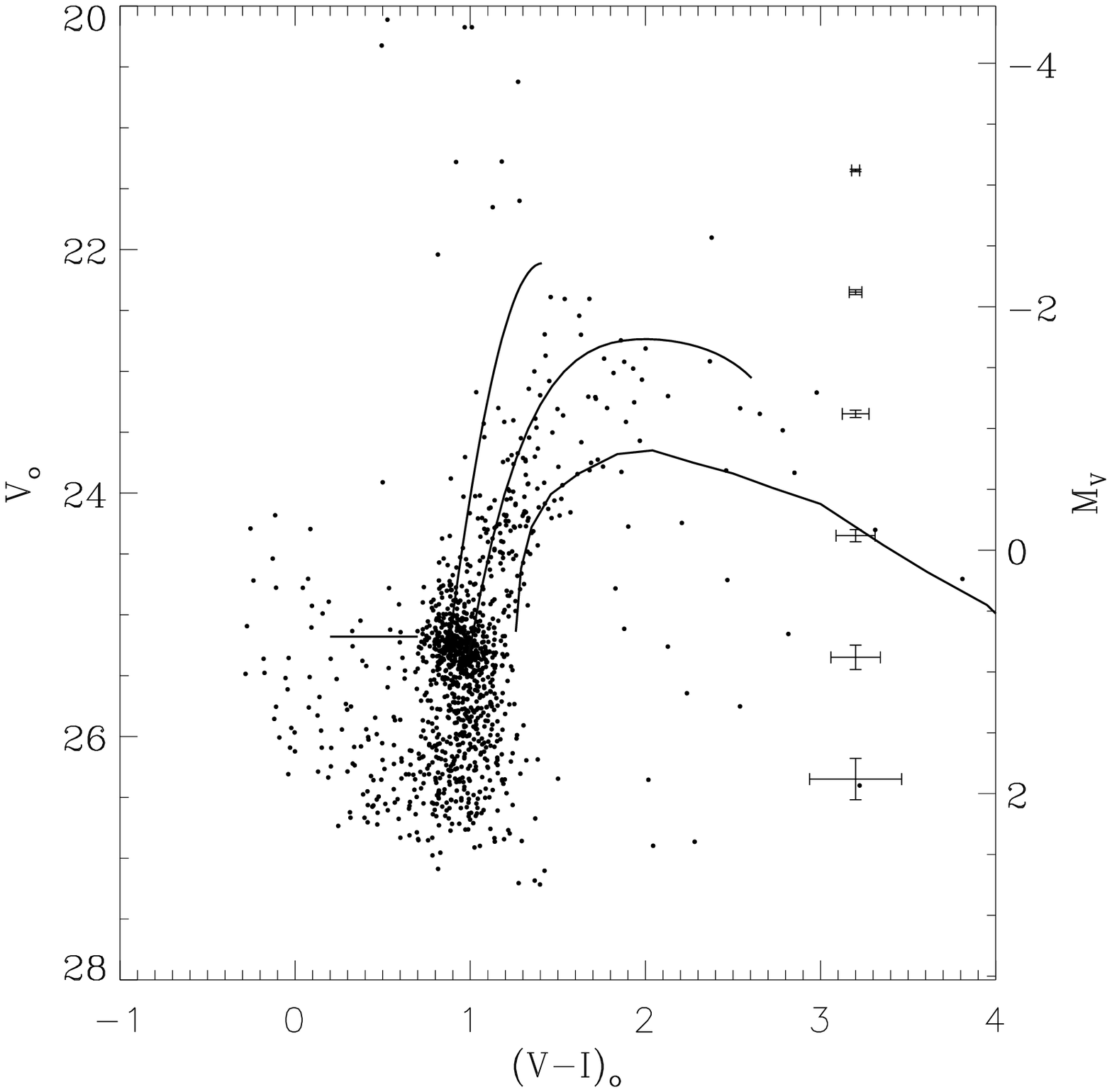,height=6cm}} {\small{\bf Fig.3 - }
(V,V$-$I) CMD for our M31 field with globular cluster fiducial sequences
overlaid.  From left to right, these correspond to the Galactic GCs
NGC~6397 ([Fe/H]$=-1.91$), 47~Tuc ([Fe/H]$=-0.71$) and NGC~6553
([Fe/H]$=-0.28$) and are taken from Da Costa \& Armandroff (1990) and
Sagar et al (1999). In the case of NGC~6553, we have assumed a distance
modulus of 13.6 and a reddening of E(V$-$I)$=0.95$ (Guarnieri et al
1998).  Also indicated is the V magnitude of the extended horizontal
branch stars detected in M31 halo fields by Holland et al (1996).}
\bigskip

We determined the mean I-band magnitude of the red clump by
constructing the luminosity function for stars with (V$-$I)$_o \ge$ 0.7
and performing a non-linear least squares fit with a function
consisting of a gaussian (to represent the RC density) and a quadratic
polynomial (to represent the background RGB density).  We calculate
I$_{RC}=24.34\pm0.05$, corresponding to M$_I=-0.13\pm0.1$, and
a fairly narrow width of $\sigma_{RC}=0.15$.  A simple gaussian fit in
color yields (V$-$I)$_{RC}=0.95$.  Assuming a metallicity of $-$0.7 dex
from the best-fitting RGB ridge line, the theoretical models of Girardi
\& Salaris (2001) indicate that a red clump this faint and this red is
most consistent with a population which has a mean age of $\gtrsim 8$
Gyr (see their Figure 1).   Younger populations (\eg  $\lesssim 4-5$
Gyr) of this metallicity would produce clumps which are $\gtrsim0.2-0.3$
mag brighter than our measurement and significantly greater than 
expected errors in our photometric zeropoint ($\sim 0.05$ mag).

Additional evidence for a predominantly old-to-intermediate age stellar
population is provided by the ratio of red clump to red giant stars,
N(RC)/N(RGB), which for  fixed metallicity and helium abundance, is
higher in younger populations.  We find  N(RC)/N(RGB) $\lesssim$ 1 in the
M31 outer disk, a
rather low value, which according to the models of Cole (1999) can be
used to exclude mean ages of $\lesssim$ 3 Gyr for the clump
population.  For ages higher than this, the variation of N(RC)/N(RGB)
with age flattens out (see also  Renzini 1994) and makes it impossible to
draw further conclusions.

There is no compelling evidence for asymptotic giant branch stars above
the tip of the red giant branch in the CMD which would represent the
luminous shell He-burning descendants of young-to-intermediate age RC
stars (2--6 Gyr). However as this evolutionary phase is extremely
rapid, such stars would need to be present in very significant numbers
to be detected given the small WFPC2 FOV (0.3 kpc$^2$ at distance of
M31).

\subsection{The Blue Population}

In addition to the dominant red population, we also detect a very
sparsely populated blue plume which extends to V$\sim 24$, V$-$I $\sim
0$ and most likely represents main sequence stars with masses in the
range $\sim 1.5-3$ {\msun}.  The existence of these more massive stars
is not surprising since our HST field lies within the outer HI disk
(Newton \& Emerson 1997) however the lack of a significant numbers of
them means that little star formation has taken place in these parts
over the last Gyr or so.

A tantalizing feature in the CMD is the population of faint stars
which appears to connect the blue plume at V$\sim 25$  to the RC.  This
feature could either represent the subgiant branch (SGB) of a $\sim1$Gyr
population, or else the extended blue horizontal branch (BHB) of ancient,
very metal poor stars ($\gtrsim 10$ Gyr, [Fe/H] $\sim -1.7$).  The
SGB interpretation has difficulties with the large number of stars
in this region compared to what would be the main sequence turnoff
region (a factor of 2:1), given that for these ages/turnoff
masses the ratio of time spent in the SGB phase relative to the
main sequence phase is of order 1\%.  On the other hand, the magnitudes
of these stars agree with the old BHB detected in
the  halo of M31 by Holland et al (1996) and indicated in Figure
3. While photometric errors and incompleteness (coupled with the
sparseness of the field) hinder deciphering the true nature of this
intriguing feature at present, we feel the available evidence best
supports its identification as a trace population of
very old, metal poor disk stars.

\section{Discussion \& Implications}

We have derived constraints on the mean age and
metallicity of stars at large radii along the major axis of M31, from
the CMD morphology of the evolved stellar populations.  The field
analysed is the only deep HST/WFPC2 pointing to date which, based on
extrapolating measured M31 structural parameters, samples the outer
disk ($\gtrsim 3$ disk scale lengths) without significant halo
contamination.

The stellar population in this field is predominantly
old-to-intermediate age (ie. $\gtrsim 8$ Gyr) with a relatively high
mean metallicity ([Fe/H]$\sim -0.7$),  indicating these stars formed
from gas which was significantly pre-enriched.  There is also a
tenative detection  of a trace population of ancient ($\ge 10$ Gyr)
metal-poor stars.   These findings are difficult to reconcile with a
scenario in which the formation of large disks is delayed to z$\lesssim 1$,  
and suggest that attempts to solve the angular
momentum problem with strong feedback are still missing an important
aspect of galaxy formation. A considerable mean age for stars at
large radii also limits the importance of late infall 
in the growth of the outer disk (eg. Ferguson \& Clarke 2001)

Additional evidence exists to support the notion that large galactic
disks have been in place for some time. This includes the
finding by Brinchmann \& Ellis (2000) that massive galaxies have formed
most of their stellar mass before a redshift of unity, and the
well-formed regular spiral galaxies at z$\sim 1$ which appear in
rest-frame optical images of the {\it Hubble Deep Field} (Ferguson,
Dickinson \& Williams 2000).  Furthermore, study of gas in the extreme
outer regions of nearby disks indicates 
prior chemical enrichment, most plausibly from previous
generations of stars in these parts (Ferguson et al 1998).

On the other hand, there are alternative interpretations of our
findings that have less serious implications for galaxy formation but
which, at the present time, appear somewhat less likely. These include:\\

{\it Disk Geometry:~~} The outer gaseous and stellar disks of M31 warp
beyond a radius of $\approx 20$kpc (Newton \& Emerson 1977, WK88), with
the disk bending northwards at large radii along the north-east major
axis.  Inspection of the deep optical plates of WK88 reveals that our
field is in the general direction of the warp and situated just below
the so-called $`$northern spur'.   Thus while it may be that our field
misses the brightest part of the stellar disk at this radius, it is
projected close enough to it that a significant fraction of disk stars
should still be detected.

{\it Disk Structure:~~}   In calculating the expected disk-to-halo
contribution at the location of the WFPC2 field, we have assumed the
disk parameters determined by WK88 can be extrapolated  beyond the
region over which they were measured (R=0--20 kpc). If the disk surface
brightness declines faster than this in the outer regions, then our
calculation will overestimate the disk fraction.  Although
several galaxies have been reported to display significant declines in
their surface brightness profiles at radii of 4--6 exponential scale
lengths (\eg van der Kruit \& Searle 1982), others continue to exhibit
exponential behaviour out to 8-10 scale lengths (\eg Weiner et al
2000). A very severe decline in disk surface brightness would be
required to render the halo stars dominant at the location of the WFPC2
field however.
 
{\it Pollution from Tidal Debris:~~} The recent discovery of a giant
tidal stream of stars near the southern minor axis of M31 indicates
that some fraction of the field halo population was not formed {\it in
situ} but was accreted from presumably smaller subsystems (Ibata et al
2001).  The similarity of the mean metallicity of stream stars, the
halo field stars and those in the far  outer disk, raises the
intriguing possibility that these stars all have a common origin.  We
cannot, at present, rule of the presence of faint halo substructure in
the vicinity of our WFPC2 field,  but note that the recently-detected
stream lies more than 120{\arcdeg} away from this location.

This {\it Letter} illustrates the potential of detailed studies of
resolved stellar populations in the local Universe  to constrain the
formation and early evolution of galaxies.   With the commissioning of
the Advanced Camera for Surveys (ACS) on HST later this year and the
advent of adaptive optics on 8-m class telescopes,  such
studies may  soon rival high redshift observations as direct tests of the
galaxy assembly process.

\acknowledgements
It is a pleasure to thank Eline Tolstoy and Andrew Cole for helpful
discussions during the course of this work, and Rosie Wyse for very
constructive comments on an early version of the manuscript.

\end{document}